\documentclass[letter, scriptaddress, twocolumn, prd, showpacs,showkeys]{revtex4-1}

	\usepackage{amsmath}
        \usepackage{mathrsfs}
	\usepackage{amssymb}
	
\makeindex

\begin{document}

\title{\large{{\bf An Exact Interior Kerr Solution}}}

\author{Aravind P Ravi}
\email{aravindravi3@gmail.com}
\author{Narayan Banerjee}
\email{narayan@iiserkol.ac.in}
\affiliation{Department of Physical Sciences, \\
Indian Institute of Science Education and Research Kolkata, \\
Mohanpur, West Bengal 741246, India.}
\begin{abstract}
 We present a simple exact solution for the interior of a rotating star. The interpretation of the stress energy tensor as that of a fluid requires the existence of a high viscosity, which is quite expected for a rotating fluid. In spite of the negative stresses, energy conditions are in fact all satisfied. 
\end{abstract}

\maketitle

A spherical symmetry is the simplest idealization of a stellar object. The exterior geometry of a static sphere is well known, which is there almost since the advent of general relativity, in the form of Schwarzschild solution\cite{schw}. An exact solution for the interior of a static perfect fluid distribution is also available, which is actually derived from the solution of pressure balance equation inside a static sphere given by Oppenheimer and Volkoff\cite{open}. The solution is a bit artificial, in the sense that it is obtained under the assumption of a constant density. Nevertheless, it is an exact solution and hence useful in many different ways. \\

In spite of its usefulness, based on the simplicity of Schwarzschild solution and its wide application potential for various estimates, the actual stars are hardly spherically symmetric. An axial symmetry is indeed a better approximation, as the stellar bodies, almost without exception, have an axial symmetry. The most useful axially symmetric exterior solution is the Kerr solution\cite{kerr} which came into being much after the discovery of the Schwarzschild solution. This solution is best expressed in the Boyer-Lindquist form\cite{boyer}, 

 \begin{equation}
 \begin{split}
 ds^{2}= dt^{2} \Bigg(1 - \frac{2mr}{\rho^{2}}\Bigg)\\ - d\phi^{2} \Bigg(\frac{2mra^{2}sin^{4}\theta}{\rho^{2}} + (r^{2} + a^{2})sin^{2}\theta\Bigg) \\
  - \rho^{2} d\theta^{2} -dr^{2} \Bigg(\frac{\rho^{2}}{r^{2}-2mr+a^{2}}\Bigg)\\
  -dt d\phi\Bigg(\frac{4mrasin^{2}\theta}{\rho^{2}}\Bigg),
  \end{split}
 \end{equation}

where $\rho^{2} = r^{2} + a^{2} cos^{2} \theta$. \\

This metric is amongst the most useful ones in general relativity as this is indeed a better description of the exterior geometry of a realistic star. However, there is hardly any interior solution for an axially symmetric star. There are definitely quite a few serious attempts. One most recent example is the one given by Hernandez-Pastora and Herrera\cite{herrera}. Their solution is based on the assumption that the exterior metric, evaluated at the boundary, enables one to extract the interior metric. The other work in the literature normally hovers around finding algorithms which might lead to an interior solution for an axially symmteric matter distribution. Newman-Janis algorithm\cite{janis} is found to be useful in this context\cite{stefano, drake}. Any such solution is normally too involved and is hardly useful for any subsequent calculation. More often than not, the  solution actually contains an arbitrary function of the radial and the polar coordinate\cite{drake}, which makes it impossible to be used for any further estimation of stellar properties.\\

In this work, we present a simple exact solution, in the Boyer-Lindquist form, for the interior of a stationary axisymmteric fluid distrbution. As no stone has perhaps been left unturned, there is actually no point in writing down the stress-energy tensor for a perfect fluid and try to solve Einstein equation for a Kerr-like metric. We adopt a very simple strategy. There are some scalar field solutions for a Kerr-like metric, quite well known in the literature. We refer to \cite{soma} for a collection of such solutions. Using an algorithm similar to that given by Eris and Gurses\cite{eris} quite a few exact scalar field solutions were derived in the work\cite{soma}. We pick up one of them, which looks favourable for the purpose of generating an interior Kerr solution, and pretend that the stress-energy tensor is that of a fluid. The rest will be a matter of interpretation and consistency checks. We show that this can be quite successfully achieved.\\

Einstein equations can be written for a minimally coupled zero mass scalar field as 
\begin{equation}\label{fe}
 R_{\mu\nu} = - (T_{\mu\nu} - \frac{1}{2} T g_{\mu\nu}) = - {\phi}_{,\mu}{\phi}_{,\nu},
\end{equation}

where $8\pi G$ is taken to be unity. The Klein Gordon equation for the scalar field, which also is a consequence of the Bianchi identities, can be written as 
\begin{equation}\label{we}
 \square \phi = 0.
\end{equation}

The chosen solution for the scalar field $\phi$ in the Boyer-Lindquist coordinates is 
\begin{equation}
 \phi = \frac{\lambda}{2}\ln [(r^2 - 2mr + a^2)\sin^{2} \theta],
\end{equation}

where $\lambda$ is a constant.

The corresponding solution for the metric is  
\begin{equation}
\begin{split}
 ds^{2} = dt^{2} - \\ 
\frac{2mr}{r^{2} + a^{2} cos^{2}\theta} (dt + a sin^{2}\theta d{\phi})^{2} \\
-(r^{2} +a^2cos^{2}\theta)[(r^{2}-2mr+a^{2})sin^{2}\theta]^{\frac{\lambda^{2}}{2}} \\ 
(d{\theta}^{2}+ \frac{dr^{2}}{r^{2}- 2mr +a^{2}}) \\
-(r^{2} +a^{2})sin^{2}\theta d{\phi}^{2}.
\end{split}
\end{equation}

Now if we pretend that the metric is due to a fluid, rather than a scalar field distribution, we do get the interior Kerr metric, where

\begin{equation}
\begin{split}
\rho = T_0^0,\\
p_r = - T_1^1, \\
p_{\theta} = - T_2^2,\\
p_{\phi} = - T_3^3,
\end{split}
\end{equation}

where the stress-energy components (using the scalar field solutions) will look like

\begin{equation}
T^{0}_{0} = A(r,\theta) \Bigg[\frac{(r-m)^{2}}{(r^{2} -2mr +a^{2})} + cot^{2} \theta \Bigg]
\end{equation}

\begin{equation}
T^{1}_{1} = A(r,\theta) \Bigg[cot^{2} \theta - \frac{(r-m)^{2}}{(r^{2} -2mr +a^{2})}  \Bigg],
\end{equation}

\begin{equation}
T^{2}_{2} =  A(r,\theta) \Bigg[-cot^{2} \theta + \frac{(r-m)^{2}}{(r^{2} -2mr +a^{2})}  \Bigg],
\end{equation}

\begin{equation}
T^{3}_{3} = A(r,\theta) \Bigg[\frac{(r-m)^{2}}{(r^{2} -2mr +a^{2})} + cot^{2} \theta \Bigg],
\end{equation}

where, $$A(r, \theta) =  \frac{\lambda^{2}}{2 (r^{2} +a^2cos^{2}\theta)[(r^{2}-2mr+a^{2})sin^{2}\theta]^{\frac{\lambda^{2}}{2}}}$$ and the constant $\lambda$ is now given by the fluid characteristics. \\

So as to keep the density positive semidefinite, one has to have $a \geq m$. This makes $p_{\phi}=-T_3^3$ a negative semidefinite quantity. Also, one has $p_r = - p_{\theta}$, so one of them is negative. The thermodynamic pressure of a real fluid cannot be negative! But a negative pressure can indeed be interpreted as a viscous stress. So one actually has an exact solution with a viscous stress. As the radial, azimuthal and the polar stresses are different, the fluid is anisotropic, which is quite an expected distribution in the absence of a spherical symmetry. \\

One can check that $\rho + p_r + p_{\theta} + p_{\phi} = 0$, which, along with $\rho \geq 0$ for $a \geq m$, ensures that the energy conditions are not violated. \\

It is important to match the interior and the exterior solutions at the boundary. We define the boundary itself as the surface where the interior and the exterior solution match. The condition yields the equation of the boundary surface as  

\begin{equation}\label{bdary}
[(r_{b}^{2}- 2mr_{b} + a^{2})sin^{2} \theta] = 1,
\end{equation}

where $r_{b}$ is the value of the radial coordinate at the boundary. At this surface, the interior and the exterior solutions do match for any arbitrary value of the fluid parameter $\lambda$. The pressure and the density of the fluid cannot extend beyond this boundary surface.  So the expressions for the fluid variables are defined from $r=0$ to the value of $r$ defined by the equation (\ref{bdary}). \\

It should be noted that for a real solution for $r_b$ for all values of $\theta$, the condition $m^2 - a^2 \geq -1$ has to be satisfied, which has no contradiction with the condition $a \geq m$ required for positive semi-definiteness of the energy density. \\

The density and pressure appear to be pathological only at $\theta = 0$ and $\theta = \pi$, i.e., along the symmetry axis. The symmetry axes are notoriously ill-behaved for an interior solution, so this pathology is not any additional deterrent. But otherwise, the solution, as well as the expressions for the density and pressure, are quite well-behaved. \\

The negative streses indeed appear to be a bit contrived, but they are quite well explained by interpreting them as viscous stresses, which is quite expected to be there in the presence of the rotation, i.e., in the absence of a spherical symmetry. The solutions are analytic and exact, and the energy conditions are found to be satisfied. It should also be recalled that the interior solution for a spherically symmetric case, much simpler to deal with,  is also based on quite an artificial assumption of a constant density.\\

The solution for the interior should match the exterior at the boundary. So the required condition $a \geq m$ should hold for the exterior solution also. It is easily seen that for $a>m$, the exterior solution is quite alright, but does not give a black hole solution as $g^{11} = 0$ yields an imaginary root for $r$ meaning that there is no event horizon. However, for $a=m$, one actually gets an event horizon, which can be called an ``extremal black hole'', as $g^{11}$ becomes a perfect square, so there is no sign flip of $g^{11}$, but the surface given by $g_{11} = 0$ indeed acts as a perfect null surface. This is somewhat analogous to an extremal Reissner-Nordstrom black hole, for which the mass of the distribution is equal to the electric charge.\\ 

Indeed there could be other solutions, with different kind of a fluid such as one with less viscosity, and even with the same distribution, as Einstein equations are nonlinear and can yield many solutions. In the absence of spherical symmetry, Birkhoff's theorem is not valid in this case, so there is no uniqueness of the solution. But in the absence of any other exact solution, the present solution should serve the purpose as there is a real requirement of the interior solution of the Kerr type metric. The interior solution presented is certainly not the one for a perfect fluid distribution, it is rather for a dissipative fluid with viscous stresses. \\

The solutions are simple, given in a closed form, and thus have the potential for being applied for purpose of investigation regarding the properties of stars as well as for the trajectories of test bodies. \\

Acknowledgement: NB thanks Naresh Dadhich for an encouraging discussion session.


\end{document}